\documentclass[10pt,conference]{IEEEtran}

\usepackage{makecell}

 \usepackage{multirow}

 \usepackage[bookmarks=false]{hyperref}
\ifCLASSINFOpdf
\else
\fi
\usepackage{relsize}

\usepackage{centernot}
\usepackage{cite}
\usepackage{amsfonts}
\usepackage{mathrsfs}
\usepackage{bbm}
\usepackage{pifont}
\usepackage{amsfonts}
\usepackage{amsmath, amsthm}
\usepackage{tabularx}
\usepackage{listings}
\usepackage{graphicx}
\usepackage{float}
\usepackage{amssymb}
\usepackage{array}
\usepackage{fancyhdr}
\usepackage{cases}
 \usepackage{supertabular}
\usepackage{url}
\usepackage{fancyhdr}

\usepackage{psfrag}
\usepackage{color}
\usepackage{bbding}
\newcommand{\bcap} {\hspace{2pt} \mathlarger{\cap}
\hspace{2pt}}

\newcommand{\acap} {\mathlarger{\cap}}

\newcommand {\C} {{\rm I\kern-5.5pt C}}

\def\centerhack#1{\hbox to 0pt{\hss\footnotesize #1\hss}}
\def\centerhackn#1{\hbox to 0pt{\hss #1\hss}}
\def\dchack#1{\vbox to 0pt{\vss{\hbox to 0pt{\hss#1\hss}}\vss}}

\newtheorem{lem}{Lemma}
\newtheorem{thm}{Theorem}

\newtheorem{rem}{Remark}

\newtheorem*{proposition1.1}{Proposition 1.1}
\newtheorem*{proposition1.2}{Proposition 1.2}
\newtheorem*{proposition1.3}{Proposition 1.3}
\newtheorem*{proposition2.1}{Proposition 2.1}
\newtheorem*{proposition2.2}{Proposition 2.2}
\renewcommand\baselinestretch{.98}

\begin{document}

\title{Random intersection graphs\\ and their applications in security, \\wireless communication, and social networks}

 \author{ \authorblockN{Jun Zhao}
\authorblockA{CyLab and Dept.
of ECE \\
Carnegie Mellon University \\ Pittsburgh, PA 15213 \\
{\tt junzhao@alumni.cmu.edu}} \and \authorblockN{Osman Ya\u{g}an}
\authorblockA{CyLab and Dept.
of ECE\\
Carnegie Mellon University \\ Moffett Field, CA 94035 \\
{\tt oyagan@ece.cmu.edu}} \and \authorblockN{Virgil Gligor}
\authorblockA{CyLab and Dept.
of ECE \\
Carnegie Mellon University \\ Pittsburgh, PA 15213 \\
{\tt gligor@cmu.edu}}}
%

\newcommand\blfootnote[1]{%
  \begingroup
  \renewcommand\thefootnote{}\footnote{#1}%
  \addtocounter{footnote}{-1}%
  \endgroup
}

\maketitle

\maketitle \thispagestyle{fancy} \pagestyle{fancy}

\fancyhead[C]{\large This invited paper in Information Theory and Applications Workshop (ITA) 2015\\summarizes some of the results in our work [40]--[54].}

\begin{abstract}

Random intersection graphs have received much interest and been used in diverse applications. They are naturally induced in modeling secure sensor networks under random key predistribution schemes, as well as in modeling the topologies of social networks including common-interest networks, collaboration networks,  and actor networks. Simply put, a random intersection graph is constructed by assigning each node a set of items in some random manner and then putting an edge between any two nodes that share a certain number of items.

Broadly speaking, our work is about analyzing random intersection graphs, and models generated by composing it with other random graph models including random geometric graphs and Erd\H{o}s--R\'enyi graphs. These compositional models are introduced to capture the characteristics of various complex natural or man-made networks more accurately than the existing models in the literature. For random intersection graphs and their compositions with other random graphs, we study properties such as ($k$-)connectivity, ($k$-)robustness, and containment of perfect matchings and Hamilton cycles. Our results are typically given in the form of asymptotically exact probabilities or zero-one laws specifying critical scalings, and provide key insights into the design and analysis of various real-world networks.


\end{abstract}

 \section{Introduction}


Random intersection graphs were introduced by Singer-Cohen
\cite{CohenThesis}. These graphs have received considerable attention in the literature [1]--[11], [28]--[35], \cite{ISIT,ZhaoAllerton,QcompTech14,ICDCSJZ,JZISIT14,ANALCO,kJZ,ZhaoCDC,2015-ISIT-phase-transition,WiOpt15,HamiltonicityJZ,2015-ISIT-s-intersection,ZhaoYaganGligor,ZhaoYaganGligor2,WiOptJZ}. In a general random intersection graph, each node is assigned a set of items in a \emph{random} manner, and any two nodes establish an \emph{undirected} edge in between if and only if they have at least a certain number of items in common. Below we explain uniform/binomial random $s$-intersection graphs that are studied in this paper.  

 In \emph{a uniform random $s$-intersection graph} with $n$ nodes, each node selects $K_n$ distinct items uniformly at random from the same item pool that has $P_n$ different items, and any two nodes have an edge in between upon sharing at least $s$ items, where $1 \leq s \leq K_n \leq P_n$ holds, and $K_n$ and $P_n$ are functions of $n$ for generality.  
  We denote a uniform random $s$-intersection graph by $G_s(n,K_n,P_n)$.  The term ``uniform'' derives from the fact that all nodes have the same number of items (but likely different sets of items).  

 In \emph{a binomial random $s$-intersection graph} with $n$ nodes, each item
from a pool of $P_n$ distinct items is assigned to each node
\emph{independently} with probability $t_n$, and any two nodes have an edge in between upon sharing at least $s$ items, where $s_n$ and $P_n$ are functions of $n$ for generality.  
  We denote a binomial random $s$-intersection graph by $H_s(n,t_n,P_n)$. The term ``binomial'' is used since the number of items
assigned to each node follows a binomial distribution with
parameters $P_n$ (the number of trials) and $t_n$ (the success
probability in each trial).

Random intersection graphs have numerous application areas including secure wireless communication \cite{ISIT,ZhaoCDC,ZhaoAllerton,JZISIT14}, social networks \cite{ball2014,mil10,DBLP:journals/corr/abs-1301-7320,virgillncs}, cryptanalysis \cite{herdingRKG}, circuit design \cite{CohenThesis}, recommender systems \cite{r4}, classification \cite{GodehardtJaworski} and clustering \cite{PES:6114960,Assortativity}. We elaborate on the use of random intersection graphs for secure wireless communication and social networks below.


\section{Use of Random Intersection Graphs for Secure Wireless Communication}

We explain below the application of random intersection graphs to secure wireless communication; 
in particular, we discuss the application of 
random intersection graphs in modeling secure wireless sensor networks.

First of all, uniform
random $1$-intersection graphs naturally capture the Eschenauer--Gligor (EG) key predistribution
scheme \cite{virgil}, which is a recognized approach to ensure secure
communications in wireless sensor networks (citation: 3700+ as of 01/07/2015). In the EG scheme for an $n$-size sensor network, 
symmetric cryptographic keys are predistributed to sensors before sensors get deployed; in particular, before deployment,
 each sensor is assigned a set of $K_n$ distinct cryptographic keys selected uniformly at random from a pool containing $P_n$ different keys. After deployment,  two sensors establish secure communication over an existing link if
and only if they have at least one common key. We say that a secure sensor network has \emph{full visibility} if secure communication between two sensors only require the key sharing and does not have link constraints (examples of link constraints include the links being reliable and the distance between sensors being small enough). Then the topology of a sensor network with the EG scheme under full visibility is given by a uniform random $1$-intersection graph $G_1(n,K_n,P_n)$.

The full visibility model explained above does not capture link constraints, but wireless links in practice might be unreliable due to the presence of
physical barriers in between or because of harsh environmental
conditions severely impairing transmission. Moreover, in real--world implementations of sensor networks, 
two sensors
have to be within a certain distance from each other to communicate, due to limited transmission ranges that result
from limited power available for transmission. Therefore, in our analysis of secure sensor networks, we consider two types of link constraints: \emph{link unreliability} and \emph{transmission constraints}. In the link unreliability model, each link between two sensors is independently active with probability $q_n$ and inactive with probability
   $(1-q_n)$. For transmission constraints, we use the widely adopted disk model: each
node's transmission area is a disk with a transmission radius $r_n$ so two nodes must have a distance at most $r_n$ for direct
communication. In terms of the node distribution, we consider that $n$ sensors
 are independently and uniformly deployed in a Euclidean plane $\mathcal {A}$, where 
$\mathcal {A}$ in our results is \emph{either} a torus $\mathcal {T}$ without any boundary or
a \emph{square} $\mathcal {S}$ with boundaries, each of a unit area.

Note that $q_n$ and $r_n$ are functions of $n$ for generality. 
The link unreliability induces an \emph{Erd\H{o}s--R\'{e}nyi graph} \cite{erdos61conn} denoted by
 $\mathcal{G}_{ER}(n,q_n)$, and the model of transmission constraints yields a \emph{random geometric graph} denoted by
$\mathcal{G}_{RGG}(n, r_n, \mathcal{A})$. In consideration of the EG scheme and the link constraints, the topology of a sensor network with the EG scheme under link unreliability is given by the intersection of a uniform random $1$-intersection graph $G_1(n,K_n,P_n)$ and an Erd\H{o}s--R\'{e}nyi graph  $\mathcal{G}_{ER}(n,q_n)$, where for graphs $G_1$ and $G_2$, two nodes have an edge in between in
$G_1\bcap G_2$ if and only if these two nodes have an edge in $G_1$
and also an edge in $G_2$. Similarly, the topology of a sensor network with the EG scheme under transmission constraints is given by the intersection of a uniform random $1$-intersection graph $G_1(n,K_n,P_n)$ and a random geometric graph
$\mathcal{G}_{RGG}(n, r_n, \mathcal{A})$.   

The EG scheme was further extended to the Chan--Perrig--Song (CPS) scheme \cite{adrian} (citation: 3000+ as of 01/07/2015). The only difference between the two schemes is that in the CPS scheme, a secure link between two sensors requires the sharing of at least $s$ different
keys rather than just one key. Then from the analysis on the EG scheme above and recalling the graph notation, 
 we immediately obtain that: (i) the topology of a sensor network with the CPS scheme under full visibility is given by 
  $G_s(n,K_n,P_n)$; (ii) the topology of a sensor network with the CPS scheme under link unreliability is given by $G_s(n,K_n,P_n) \bcap \mathcal{G}_{ER}(n,q_n)$; and (iii) the topology of a sensor network with the CPS scheme under transmission constraints is given by $G_s(n,K_n,P_n) \bcap \mathcal{G}_{RGG}(n, r_n, \mathcal{A})$.
  
  Table \ref{table:wsn} summarizes different settings of secure sensor networks and their corresponding random graph models.

 \begin{table}[t]
\begin{tabular}{|l|l|l|}
\hline
\multicolumn{2}{|l|}{\hspace{-3pt}Settings\hspace{-3pt}}                         & \hspace{-3pt}Graphs\hspace{-3pt}           \\ \hline
\multirow{3}{*}{\hspace{-3pt}EG scheme\hspace{-3pt}}  & \hspace{-3pt}full visibility\hspace{-3pt}          & \hspace{-3pt}$G_1(n,K_n,P_n)$\hspace{-3pt} \\ \cline{2-3} 
                            & \hspace{-3pt}link unreliability\hspace{-3pt}       & \hspace{-3pt}$G_1(n,K_n,P_n) \bcap \mathcal{G}_{ER}(n,q_n)$\hspace{-3pt}                \\ \cline{2-3} 
                            & \hspace{-3pt}transmission constraints\hspace{-3pt} & \hspace{-3pt}$G_1(n,K_n,P_n) \bcap \mathcal{G}_{RGG}(n, r_n, \mathcal{A})$\hspace{-3pt}                \\ \hline
\multirow{3}{*}{\hspace{-3pt}CPS scheme\hspace{-3pt}} & \hspace{-3pt}full visibility\hspace{-3pt}          & \hspace{-3pt}$G_s(n,K_n,P_n)$\hspace{-3pt} \\ \cline{2-3} 
                            & \hspace{-3pt}link unreliability\hspace{-3pt}       & \hspace{-3pt}$G_s(n,K_n,P_n) \bcap \mathcal{G}_{ER}(n,q_n)$\hspace{-3pt}                \\ \cline{2-3} 
                            & \hspace{-3pt}transmission constraints\hspace{-3pt} & \hspace{-3pt}$G_s(n,K_n,P_n) \bcap \mathcal{G}_{RGG}(n, r_n, \mathcal{A})$\hspace{-3pt}                \\ \hline
\end{tabular}
  \caption{Different settings of secure sensor networks and their corresponding random graph models.\vspace{-10pt}}\vspace{-13pt}
  \label{table:wsn}
\end{table}

\section{Use of Random Intersection Graphs for Social Networks}

We explain 
that random intersection graphs are natural models for social networks \cite{RIGbook}, examples of which given below are common-interest networks, researcher networks and actor networks. In a
common-interest networks \cite{ZhaoYaganGligor},
each user has several interests following some distribution, and two users are said to have a 
common-interest relation if they share at least $s$ interest(s). In a 
researcher network (an example of a collaboration network) \cite{bloznelis2013,PES:6114960}, each researcher publishes a number of papers, and two researchers are adjacent if co-authoring at least s paper(s). In an actor network 
\cite{DBLP:journals/corr/abs-1301-7320,mil10}, each actor contributes to a number of films, and
two actors are adjacent if acting in at least s common film(s). Examples can be extended to other 
types of social networks. For all examples, clearly the induced topologies are represented by random intersection graphs.

\section{A Summary of Results} \label{sec:main:res}

We present below the results for random intersection graphs, and their compositions with other random graphs in terms of various properties including $k$-connectivity, perfect matching containment, Hamilton cycle containment, and $k$-robustness. These properties are defined as follows: (i) A graph is $k$-connected if each pair of nodes has at least $k$
internally node-disjoint path(s) between them, and a graph is connected if it is $1$-connected.
 (ii)
A perfect matching is a set of edges that do not have common nodes
and cover all nodes with the exception of missing at most one node. (iii) A Hamiltonian cycle is a closed loop that visits each node exactly once. (iv)
The notion of $k$-robustness proposed by Zhang and
Sundaram \cite{6425841} measures the effectiveness of local-information-based diffusion algorithms in the presence of adversarial nodes; formally, a graph with a node set $\mathcal {V}$ is $k$-robust
 if at least one of (a) and (b) below holds for each
non-empty and strict subset $T$ of $\mathcal {V}$: (a) there exists
at least a node $v_a \in T$ such that
  $v_a$ has no less than $k$ neighbors inside $\mathcal {V}\setminus
  T$, and (b) there exists at least a node $v_b \in \mathcal {V}\setminus T$ such that
  $v_b$ has no less than $k$ neighbors inside $T$, where two nodes are neighbors if they have an edge in between. This notion of $k$-robustness has received much attention
\cite{6481629,leblanc2013resilient,zhang2012robustness,zhang2012simple,ZhaoCDC,2015-ISIT-s-intersection}.

Notation and convention: Throughout the paper, both $k$ and $s$ are positive constant integers and do not scale with $n$.  All asymptotic statements are understood with $ n \to \infty$. We use the  Landau asymptotic notation $O(\cdot), o(\cdot), \Omega(\cdot),
\omega(\cdot), \Theta(\cdot), \sim$; in particular, for two
positive sequences $x_n$ and $y_n$, the relation $x_n \sim y_n$
signifies $\lim_{n \to
  \infty} (x_n/y_n)=1$. Also, $\mathbb{P}[\mathcal {E}]$
denotes the probability that event $\mathcal {E}$ occurs. An event happens \emph{almost surely} if its probability converges to $1$ as $n\to\infty$.

\subsection{Results of random intersection graphs}  

\subsubsection{Results of uniform random $1$-intersection graphs}

\begin{thm}[\hspace{0pt}{\textbf{$k$-Connectivity in uniform random $1$-intersection graphs by our work \cite{ZhaoCDC}}}\hspace{0pt}]   \label{thm-a1}
 
For a uniform random $1$-intersection graph
$G_1(n,K_n,P_n)$, if there is a sequence $\alpha_n$ with $\lim_{n \to \infty}{\alpha_n} \in [-\infty, \infty]$
such that
\begin{align}
 \frac{{K_n}^2}{P_n} & = \frac{\ln  n   + {(k-1)} \ln \ln n  +
 {\alpha_n}}{n}, \nonumber
\end{align}
then under $P_n = \Omega(n)$, it holds that\begin{align}
&  \lim_{n \to \infty}\mathbb{P} \left[\hspace{2pt}G_1(n,K_n,P_n)\textrm{
is $k$-connected}.\hspace{2pt}\right] \nonumber \\
 & \hspace{-2pt}=\hspace{-2pt}  \lim_{n \to \infty}\mathbb{P} \left[\hspace{2pt}G_1(n,K_n,P_n)\textrm{
has a minimum degree at least $k$}.\hspace{2pt}\right]  \nonumber \\ &    \hspace{-2pt}=\hspace{-2pt} e^{- \frac{e^{-\lim_{n \to \infty}{\alpha_n}}}{(k-1)!}} \hspace{-2pt}=\hspace{-2pt}  \begin{cases} 0,&\text{\hspace{-8pt}if  }\lim_{n \to \infty}{\alpha_n} \hspace{-2pt}=\hspace{-2pt}- \infty, \\
1,&\text{\hspace{-8pt}if  }\lim_{n \to \infty}{\alpha_n} \hspace{-2pt}=\hspace{-2pt} \infty, \\
e^{- \frac{e^{-\alpha^{*}}}{(k-1)!}},&\text{\hspace{-8pt}if  }\lim_{n \to \infty}{\alpha_n} \hspace{-2pt}=\hspace{-2pt} \alpha^{*}\hspace{-2pt}\in\hspace{-2pt} (-\infty, \hspace{-1.5pt}\infty). \end{cases}\nonumber
 \end{align}

\end{thm}

\begin{rem}

Theorem \ref{thm-a1} presents the asymptotically exact probability of $k$-connectivity in a uniform random $1$-intersection graph, while 
 a zero--one law  is  implicitly obtained by Rybarczyk \cite{zz} and  
explicitly given by us as a side result \cite{ZhaoYaganGligor,ISIT}. 
For connectivity (i.e., $k$-connectivity in the case of $k=1$), Blackburn and Gerke \cite{r1} and Ya\u{g}an and Makowski \cite{yagan} show different granularities of zero--one laws, while  Rybarczyk \cite{ryb3} derives
the asymptotically exact probability.
\end{rem}

\begin{thm}[\hspace{0pt}{\textbf{Perfect matching containment in uniform random $1$-intersection graphs by our work \cite{2015-ISIT-phase-transition}}}\hspace{0pt}]   \label{thm-a2}

For a uniform random $1$-intersection graph
$G_1(n,K_n,P_n)$, if there is a sequence $\beta_n$ with $\lim_{n \to \infty}{\beta_n} \in [-\infty, \infty]$
such that
\begin{align}
 \frac{{K_n}^2}{P_n} & = \frac{\ln  n   +
 {\beta_n}}{n}, \nonumber
\end{align}
then under $P_n = \omega\big(n (\ln n)^5\big)$, it holds that
 \begin{align}
& \lim_{n \to \infty}  \mathbb{P}[\hspace{2pt}G_1(n, K_n, P_n)\text{ has at least one perfect matching.}\hspace{2pt}] \nonumber \\ &   \hspace{-1pt}=\hspace{-1.5pt} e^{- e^{-\lim\limits_{n \to \infty}{\beta_n}}} \hspace{-1.5pt}=\hspace{-1.5pt}  \begin{cases} 0,&\text{\hspace{-4pt}if  }\lim_{n \to \infty}{\beta_n} \hspace{-1.5pt}=\hspace{-1.5pt}- \infty, \\
1,&\text{\hspace{-4pt}if  }\lim_{n \to \infty}{\beta_n} \hspace{-1.5pt}=\hspace{-1.5pt} \infty, \\
e^{-e^{- \beta^{*}}},&\text{\hspace{-4pt}if  }\lim_{n \to \infty}{\beta_n} \hspace{-1.5pt}=\hspace{-1.5pt} \beta^{*}\hspace{-1.5pt}\in\hspace{-1.5pt} (-\infty, \hspace{-1.5pt}\infty). \end{cases}\nonumber
 \end{align}
\end{thm}

\begin{rem}

Theorem \ref{thm-a2} presents the asymptotically exact probability of perfect matching containment in a uniform random $1$-intersection graph. A similar result is given by setting $s$ as $1$ in the work of Bloznelis and {\L}uczak \cite{Perfectmatchings} studying $G_s(n,K_n,P_n)$. However, they use conditions $K_n = O\big((\ln n)^{\frac{1}{5}}\big)$ and $ \frac{{K_n}^2}{P_n} = O\big(\frac{\ln  n }{n}\big)$. Furtherermore, for the one-law (i.e., the case where $G_1(n, K_n, P_n)$ contains a perfect matching almost surely), their result relies on $P_n = o\big(n (\ln n)^{-\frac{3}{5}}\big)$, whereas our result uses $P_n = \omega\big(n (\ln n)^5\big)$. We note that $P_n$ is expected to be at least on the order of $n$
in the sensor network applications of uniform random $1$-intersection graphs  \cite{virgil}. In addition, Blackburn \emph{et al.} \cite{r10} derive a result that is weaker than Theorem \ref{thm-a2}, to analyze cryptographic hash functions. Specifically, they show that for a uniform random $1$-intersection graph
$G_1(n,K_n,P_n)$ under $P_n = \Omega(n^c)$ with a constant $c>1$, then $G_1(n,K_n,P_n)$ contains (resp., does not contain) a perfect matching almost surely if $\lim_{n \to \infty} \big(\frac{{K_n}^2}{P_n} \big/ \frac{\ln n}{n} \big) > 1$ (resp., $<1$).
 
\end{rem}

\begin{thm}[\hspace{0pt}{\textbf{Hamilton cycle containment in uniform random $1$-intersection graphs by our work \cite{2015-ISIT-phase-transition}}}\hspace{0pt}]   \label{thm:HC}

For a uniform random $1$-intersection graph
$G_1(n,K_n,P_n)$, if there is a sequence $\gamma_n$ with $\lim_{n \to \infty}{\gamma_n} \in [-\infty, \infty]$
such that
\begin{align}
 \frac{{K_n}^2}{P_n} & = \frac{\ln  n +  \ln \ln n +
 {\gamma_n}}{n}, \nonumber
\end{align}
then under $P_n = \omega\big(n (\ln n)^5\big)$, it holds that
 \begin{align}
& \lim_{n \to \infty}  \mathbb{P}[\hspace{2pt}G_1(n, K_n, P_n)\text{ has at least one Hamilton cycle.}\hspace{2pt}] \nonumber \\ &   \hspace{-1pt}=\hspace{-1.5pt} e^{- e^{-\lim\limits_{n \to \infty}{\gamma_n}}} \hspace{-1.5pt}=\hspace{-1.5pt}  \begin{cases} 0,&\text{\hspace{-4pt}if  }\lim_{n \to \infty}{\gamma_n} \hspace{-1.5pt}=\hspace{-1.5pt}- \infty, \\
1,&\text{\hspace{-4pt}if  }\lim_{n \to \infty}{\gamma_n} \hspace{-1.5pt}=\hspace{-1.5pt} \infty, \\
e^{-e^{- \gamma^{*}}},&\text{\hspace{-4pt}if  }\lim_{n \to \infty}{\gamma_n} \hspace{-1.5pt}=\hspace{-1.5pt} \gamma^{*}\hspace{-1.5pt}\in\hspace{-1.5pt} (-\infty, \hspace{-1.5pt}\infty). \end{cases}\nonumber
 \end{align}
\end{thm}
 
\begin{rem}

Nikoletseas \emph{et al. } \cite{NikoletseasHM} proves that 
 $G_1(n, K_n, P_n)$ under $K_n \geq 2$ has a
    Hamilton cycle with high probability 
     if it holds for some constant $\delta>0$ that $n \geq (1+\delta) \binom{P_n}{K_n} \ln \binom{P_n}{K_n} $, which implies that $P_n$ is much smaller than $n$ ($P_n = O(\sqrt{n}\hspace{1.5pt})$ given $K_n \geq 2$,
    $P_n = O(\sqrt[3]{n}\hspace{1.5pt})$ if $K_n \geq 3$, $P_n = O(\sqrt[4]{n}\hspace{1.5pt})$ if $K_n \geq 4$, etc.). Different from the result of Nikoletseas \emph{et al. } \cite{NikoletseasHM}, our Theorem \ref{thm:HC} is for $P_n = \omega\big(n (\ln n)^5\big)$. Furthermore, Theorem \ref{thm:HC} presents the asymptotically exact probability, whereas Nikoletseas \emph{et al. } \cite{NikoletseasHM} only derive conditions for $G_1(n, K_n, P_n)$ to have a Hamilton cycle almost surely. They do not provide conditions for $G_1(n, K_n, P_n)$ to have no Hamilton cycle with high probability, or to have a Hamilton cycle with an asymptotic probability in $(0,1)$.
 
\end{rem}

\begin{thm}[\hspace{0pt}{\textbf{$k$-Robustness in uniform random $1$-intersection graphs by our work \cite{ZhaoCDC}}}\hspace{0pt}]   
 
For a uniform random $1$-intersection graph
$G_1(n,K_n,P_n)$, with a sequence $\delta_n$ defined
by
\begin{align}
 \frac{{K_n}^2}{P_n} & = \frac{\ln  n   + {(k-1)} \ln \ln n  +
 {\delta_n}}{n}, \nonumber
\end{align}
then under $P_n = \Omega\big(n (\ln n)^5\big)$, it holds that
\begin{align}
&  \lim_{n \to \infty}\mathbb{P} \left[\hspace{2pt}G_1(n,K_n,P_n)\textrm{
is $k$-robust}.\hspace{2pt}\right]   \nonumber \\ & \quad =   \begin{cases} 0,&\text{if  }\lim_{n \to \infty}{\delta_n}  = - \infty, \\
1,&\text{if  }\lim_{n \to \infty}{\delta_n}  =  \infty. \end{cases}\nonumber
 \end{align}

\end{thm}

\begin{rem}
As mentioned earlier, we use the definition of $k$-robustness proposed by Zhang and
Sundaram \cite{6425841}. They present results on  $k$-robustness in Erd\H{o}s--R\'{e}nyi graphs and one-dimensional random geometric graphs, whereas we study their notion of $k$-robustness in random intersection graphs \cite{ZhaoCDC,2015-ISIT-s-intersection}.
\end{rem}

\subsubsection{Results of binomial random $1$-intersection graphs}

\begin{thm}[\hspace{0pt}{\textbf{$k$-Connectivity in binomial random $1$-intersection graphs by our work \cite{ZhaoCDC}}}\hspace{0pt}]   \label{thm-a1-2}
 
For a binomial random $1$-intersection graph
$H_1(n,t_n,P_n)$, if there is a sequence $\alpha_n$ with $\lim_{n \to \infty}{\alpha_n} \in [-\infty, \infty]$
such that
\begin{align}
 {t_n}^2 {P_n} & = \frac{\ln  n   + {(k-1)} \ln \ln n  +
 {\alpha_n}}{n}, \nonumber
\end{align}
then under $P_n = \omega\big(n (\ln n)^5\big)$, it holds that\begin{align}
&  \lim_{n \to \infty}\mathbb{P} \left[\hspace{2pt}H_1(n,t_n,P_n)\textrm{
is $k$-connected}.\hspace{2pt}\right] \nonumber \\
 & \hspace{-2pt}=\hspace{-2pt}  \lim_{n \to \infty}\mathbb{P} \left[\hspace{2pt}H_1(n,t_n,P_n)\textrm{
has a minimum degree at least $k$}.\hspace{2pt}\right]  \nonumber \\ &    \hspace{-2pt}=\hspace{-2pt} e^{- \frac{e^{-\lim_{n \to \infty}{\alpha_n}}}{(k-1)!}} \hspace{-2pt}=\hspace{-2pt}  \begin{cases} 0,&\text{\hspace{-8pt}if  }\lim_{n \to \infty}{\alpha_n} \hspace{-2pt}=\hspace{-2pt}- \infty, \\
1,&\text{\hspace{-8pt}if  }\lim_{n \to \infty}{\alpha_n} \hspace{-2pt}=\hspace{-2pt} \infty, \\
e^{- \frac{e^{-\alpha^{*}}}{(k-1)!}},&\text{\hspace{-8pt}if  }\lim_{n \to \infty}{\alpha_n} \hspace{-2pt}=\hspace{-2pt} \alpha^{*}\hspace{-2pt}\in\hspace{-2pt} (-\infty, \hspace{-1.5pt}\infty). \end{cases}\nonumber
 \end{align}

\end{thm}

\begin{rem}

Theorem \ref{thm-a1-2} presents the asymptotically exact probability of $k$-connectivity in a binomial random $1$-intersection graph, while zero--one laws are obtained by Rybarczyk \cite{zz,2013arXiv1301.0466R}. 
Connectivity (i.e., $k$-connectivity in the case of $k=1$) results are presented by Rybarczyk \cite{2013arXiv1301.0466R,zz}, 
Shang \cite{YShang}, and
 Singer-Cohen \cite{CohenThesis}.

\end{rem}

\begin{thm}[\hspace{0pt}{\textbf{Perfect matching containment in binomial random $1$-intersection graphs by Rybarczyk  
\cite{zz,2013arXiv1301.0466R}}}\hspace{0pt}]   \label{pmc}

For a binomial random $1$-intersection graph
$H_1(n,t_n,P_n)$, if there is a sequence $\beta_n$ with $\lim_{n \to \infty}{\beta_n} \in [-\infty, \infty]$
such that
\begin{align}
 {t_n}^2 {P_n} & = \frac{\ln  n   +
 {\beta_n}}{n}, \label{tnpnn}
\end{align}
then under $P_n = \Omega(n^c)$ for a constant $c>1$, it holds that
 \begin{align}
& \lim_{n \to \infty}  \mathbb{P}[\hspace{2pt}H_1(n, t_n, P_n)\text{ has at least one perfect matching.}\hspace{2pt}] \nonumber \\ &   \hspace{-1pt}=\hspace{-1.5pt} e^{- e^{-\lim\limits_{n \to \infty}{\beta_n}}} \hspace{-1.5pt}=\hspace{-1.5pt}  \begin{cases} 0,&\text{\hspace{-4pt}if  }\lim_{n \to \infty}{\beta_n} \hspace{-1.5pt}=\hspace{-1.5pt}- \infty, \\
1,&\text{\hspace{-4pt}if  }\lim_{n \to \infty}{\beta_n} \hspace{-1.5pt}=\hspace{-1.5pt} \infty, \\
e^{-e^{- \beta^{*}}},&\text{\hspace{-4pt}if  }\lim_{n \to \infty}{\beta_n} \hspace{-1.5pt}=\hspace{-1.5pt} \beta^{*}\hspace{-1.5pt}\in\hspace{-1.5pt} (-\infty, \hspace{-1.5pt}\infty). \end{cases}\nonumber
 \end{align}
\end{thm}
 
\begin{rem}
For perfect matching containment in a binomial random $1$-intersection graph, in addition to Theorem \ref{pmc} above under $P_n = \Omega(n^c)$ for a constant $c>1$,
Rybarczyk  
\cite{zz,2013arXiv1301.0466R} also derives results under $P_n = \Omega(n^c)$ for a constant $c<1$, with a scaling condition different from (\ref{tnpnn}).
 
\end{rem}

\begin{thm}[\hspace{0pt}{\textbf{Hamilton cycle containment in binomial random $1$-intersection graphs by our work \cite{HamiltonicityJZ}}}\hspace{0pt}]   \label{thm-a1-3} 

For a binomial random $1$-intersection graph
$H_1(n,t_n,P_n)$, if there is a sequence $\gamma_n$ with $\lim_{n \to \infty}{\gamma_n} \in [-\infty, \infty]$
such that
\begin{align}
 {t_n}^2 {P_n} & = \frac{\ln  n +  \ln \ln n +
 {\gamma_n}}{n}, \nonumber
\end{align}
then under $P_n = \omega\big(n (\ln n)^5\big)$, it holds that
 \begin{align}
& \lim_{n \to \infty}  \mathbb{P}[\hspace{2pt}H_1(n, t_n, P_n)\text{ has at least one Hamilton cycle.}\hspace{2pt}] \nonumber \\ &   \hspace{-1pt}=\hspace{-1.5pt} e^{- e^{-\lim\limits_{n \to \infty}{\gamma_n}}} \hspace{-1.5pt}=\hspace{-1.5pt}  \begin{cases} 0,&\text{\hspace{-4pt}if  }\lim_{n \to \infty}{\gamma_n} \hspace{-1.5pt}=\hspace{-1.5pt}- \infty, \\
1,&\text{\hspace{-4pt}if  }\lim_{n \to \infty}{\gamma_n} \hspace{-1.5pt}=\hspace{-1.5pt} \infty, \\
e^{-e^{- \gamma^{*}}},&\text{\hspace{-4pt}if  }\lim_{n \to \infty}{\gamma_n} \hspace{-1.5pt}=\hspace{-1.5pt} \gamma^{*}\hspace{-1.5pt}\in\hspace{-1.5pt} (-\infty, \hspace{-1.5pt}\infty). \end{cases}\nonumber
 \end{align}
\end{thm}

\begin{rem}

Theorem \ref{thm-a1-3} presents the asymptotically exact probability of Hamilton cycle containment in a binomial random $1$-intersection graph, while zero--one laws are obtained by Efthymioua and Spirakis \cite{EfthymiouaHM}, and Rybarczyk \cite{zz,2013arXiv1301.0466R}.
\end{rem}

\begin{thm}[\hspace{0pt}{\textbf{$k$-Robustness in binomial random $1$-intersection graphs by our work \cite{ZhaoCDC}}}\hspace{0pt}]   
 
For a binomial random $1$-intersection graph
$H_1(n,t_n,P_n)$, with a sequence $\delta_n$ defined
by
\begin{align}
 {t_n}^2 {P_n} & = \frac{\ln  n   + {(k-1)} \ln \ln n  +
 {\delta_n}}{n}, \nonumber
\end{align}
then under $P_n = \Omega\big(n (\ln n)^5\big)$, it holds that  
 \begin{align}
&  \lim_{n \to \infty}\mathbb{P} \left[\hspace{2pt}H_1(n,t_n,P_n)\textrm{
is $k$-robust}.\hspace{2pt}\right]   \nonumber \\ & \quad =   \begin{cases} 0,&\text{if  }\lim_{n \to \infty}{\delta_n}  = - \infty, \\
1,&\text{if  }\lim_{n \to \infty}{\delta_n}  =  \infty. \end{cases}\nonumber
 \end{align}

\end{thm}

\subsubsection{Results of uniform random $s$-intersection graphs}

\begin{thm}[\hspace{0pt}{\textbf{$k$-Connectivity in uniform random $s$-intersection graphs by our work \cite{ANALCO}}}\hspace{0pt}]  \label{thm-a1-5} 

For a uniform random $s$-intersection graph
$G_s(n,K_n,P_n)$, if there is a sequence $\alpha_n$ with $\lim_{n \to \infty}{\alpha_n} \in [-\infty, \infty]$
such that
\begin{align}
\frac{1}{s!} \cdot \frac{{K_n}^{2s}}{{P_n}^{s}}& = \frac{\ln  n   + {(k-1)} \ln \ln n  +
 {\alpha_n}}{n}, \nonumber
\end{align}
then under $P_n  =  \Omega(n^c)$ for
a constant $c>2-\frac{1}{s}$, it holds that
\begin{align}
&  \lim_{n \to \infty}\mathbb{P} \left[\hspace{2pt}G_s(n,K_n,P_n)\textrm{
is $k$-connected}.\hspace{2pt}\right] \nonumber \\
 & \hspace{-2pt}=\hspace{-2pt}  \lim_{n \to \infty}\mathbb{P} \left[\hspace{2pt}G_s(n,K_n,P_n)\textrm{
has a minimum degree at least $k$}.\hspace{2pt}\right]  \nonumber \\ &    \hspace{-2pt}=\hspace{-2pt} e^{- \frac{e^{-\lim_{n \to \infty}{\alpha_n}}}{(k-1)!}} \hspace{-2pt}=\hspace{-2pt}  \begin{cases} 0,&\text{\hspace{-8pt}if  }\lim_{n \to \infty}{\alpha_n} \hspace{-2pt}=\hspace{-2pt}- \infty, \\
1,&\text{\hspace{-8pt}if  }\lim_{n \to \infty}{\alpha_n} \hspace{-2pt}=\hspace{-2pt} \infty, \\
e^{- \frac{e^{-\alpha^{*}}}{(k-1)!}},&\text{\hspace{-8pt}if  }\lim_{n \to \infty}{\alpha_n} \hspace{-2pt}=\hspace{-2pt} \alpha^{*}\hspace{-2pt}\in\hspace{-2pt} (-\infty, \hspace{-1.5pt}\infty). \end{cases}\nonumber
 \end{align}

\end{thm}

\begin{rem}

Theorem \ref{thm-a1-5} presents the asymptotically exact probability of $k$-connectivity in a uniform random $s$-intersection graph, while 
a similar result for $k$-connectivity is given by Bloznelis and Rybarczyk \cite{Bloznelis201494}, and a similar result for connectivity (i.e., $k$-connectivity in the case of $k=1$) is shown by Bloznelis and {\L}uczak \cite{Perfectmatchings}, but both results \cite{Bloznelis201494,Perfectmatchings} assume $K_n = O\big((\ln n)^{\frac{1}{5s}}\big)$, which limits their applications to secure sensor networks \cite{adrian}. 
\end{rem}

\begin{thm}[\hspace{0pt}{\textbf{Perfect matching containment in uniform random $s$-intersection graphs by our work \cite{2015-ISIT-s-intersection}}}\hspace{0pt}]   \label{pm:thm:unixx}
For a uniform random $s$-intersection graph
$G_s(n,K_n,P_n)$, if there is a sequence $\beta_n$ with $\lim_{n \to \infty}{\beta_n} \in [-\infty, \infty]$
such that
\begin{align}
\frac{1}{s!} \cdot \frac{{K_n}^{2s}}{{P_n}^{s}}& = \frac{\ln  n  +
 {\beta_n}}{n}, \nonumber
\end{align}
then under $P_n  =  \Omega(n^c)$ for
a constant $c>2-\frac{1}{s}$, it holds that
  \begin{align}
& \lim_{n \to \infty}  \mathbb{P}[\hspace{2pt}G_s(n,K_n,P_n)\text{ has at least one perfect matching.}\hspace{2pt}] \nonumber \\ &   \hspace{-1pt}=\hspace{-1.5pt} e^{- e^{-\lim\limits_{n \to \infty}{\beta_n}}} \hspace{-1.5pt}=\hspace{-1.5pt}  \begin{cases} 0,&\text{\hspace{-4pt}if  }\lim_{n \to \infty}{\beta_n} \hspace{-1.5pt}=\hspace{-1.5pt}- \infty, \\
1,&\text{\hspace{-4pt}if  }\lim_{n \to \infty}{\beta_n} \hspace{-1.5pt}=\hspace{-1.5pt} \infty, \\
e^{-e^{- \beta^{*}}},&\text{\hspace{-4pt}if  }\lim_{n \to \infty}{\beta_n} \hspace{-1.5pt}=\hspace{-1.5pt} \beta^{*}\hspace{-1.5pt}\in\hspace{-1.5pt} (-\infty, \hspace{-1.5pt}\infty). \end{cases}\nonumber
 \end{align}
\end{thm}

\begin{rem}

Theorem \ref{pm:thm:unixx} presents the asymptotically exact probability of perfect matching containment  in a uniform random $s$-intersection graph, while 
a similar result is given by  Bloznelis and {\L}uczak \cite{Perfectmatchings} under $K_n = O\big((\ln n)^{\frac{1}{5s}}\big)$. 
\end{rem}

\begin{thm}[\hspace{0pt}{\textbf{Hamilton cycle containment in uniform random $s$-intersection graphs by our work \cite{2015-ISIT-s-intersection}}}\hspace{0pt}]    \label{hc:thm:uni}
For a uniform random $s$-intersection graph
$G_s(n,K_n,P_n)$, if there is a sequence $\gamma_n$ with $\lim_{n \to \infty}{\gamma_n} \in [-\infty, \infty]$
such that
\begin{align}
\frac{1}{s!} \cdot \frac{{K_n}^{2s}}{{P_n}^{s}}& = \frac{\ln  n   + \ln \ln n  +
 {\gamma_n}}{n}, \nonumber
\end{align}
then under $P_n  =  \Omega(n^c)$ for
a constant $c>2-\frac{1}{s}$, it holds that
 \begin{align}
& \lim_{n \to \infty}  \mathbb{P}[\hspace{2pt}G_s(n, K_n, P_n)\text{ has at least one Hamilton cycle.}\hspace{2pt}] \nonumber \\ &   \hspace{-1pt}=\hspace{-1.5pt} e^{- e^{-\lim\limits_{n \to \infty}{\gamma_n}}} \hspace{-1.5pt}=\hspace{-1.5pt}  \begin{cases} 0,&\text{\hspace{-4pt}if  }\lim_{n \to \infty}{\gamma_n} \hspace{-1.5pt}=\hspace{-1.5pt}- \infty, \\
1,&\text{\hspace{-4pt}if  }\lim_{n \to \infty}{\gamma_n} \hspace{-1.5pt}=\hspace{-1.5pt} \infty, \\
e^{-e^{- \gamma^{*}}},&\text{\hspace{-4pt}if  }\lim_{n \to \infty}{\gamma_n} \hspace{-1.5pt}=\hspace{-1.5pt} \gamma^{*}\hspace{-1.5pt}\in\hspace{-1.5pt} (-\infty, \hspace{-1.5pt}\infty). \end{cases}\nonumber
 \end{align}
\end{thm}

\begin{thm}[\hspace{0pt}{\textbf{$k$-Robustness in uniform random $s$-intersection graphs by our work \cite{2015-ISIT-s-intersection}}}\hspace{0pt}]     \label{rb:thm:uni}
For a uniform random $s$-intersection graph
$G_s(n,K_n,P_n)$, with a sequence $\delta_n$ defined
by
\begin{align}
\frac{1}{s!} \cdot \frac{{K_n}^{2s}}{{P_n}^{s}}& = \frac{\ln  n   + {(k-1)} \ln \ln n  +
 {\delta_n}}{n}, \nonumber
\end{align}
then under $P_n  =  \Omega(n^c)$ for
a constant $c>2-\frac{1}{s}$, it holds that
  \begin{align}
&  \lim_{n \to \infty}\mathbb{P} \left[\hspace{2pt}G_s(n,K_n,P_n)\textrm{
is $k$-robust}.\hspace{2pt}\right]   \nonumber \\ & \quad =   \begin{cases} 0,&\text{if  }\lim_{n \to \infty}{\delta_n}  = - \infty, \\
1,&\text{if  }\lim_{n \to \infty}{\delta_n}  =  \infty. \end{cases}\nonumber
 \end{align}
\end{thm}

\subsubsection{Results of binomial random $s$-intersection graphs}

\begin{thm}[\hspace{0pt}{\textbf{$k$-Connectivity in binomial random $s$-intersection graphs by our work \cite{ANALCO}}}\hspace{0pt}]  
  \label{kcon:lem:bin}
For a binomial random $s$-intersection graph
$H_s(n,t_n,P_n)$, if there is a sequence $\alpha_n$ with $\lim_{n \to \infty}{\alpha_n} \in [-\infty, \infty]$
such that
\begin{align}
\frac{1}{s!} \cdot {t_n}^{2s}{P_n}^{s} & = \frac{\ln  n   + {(k-1)} \ln \ln n  +
 {\alpha_n}}{n}, \nonumber
\end{align}
then under $P_n  =  \Omega(n^c)$ for
a constant $c>2-\frac{1}{s}$, it holds that
\begin{align}
&  \lim_{n \to \infty}\mathbb{P} \left[\hspace{2pt}H_s(n,t_n,P_n)\textrm{
is $k$-connected}.\hspace{2pt}\right] \nonumber \\
 & \hspace{-2pt}=\hspace{-2pt}  \lim_{n \to \infty}\mathbb{P} \left[\hspace{2pt}H_s(n,t_n,P_n)\textrm{
has a minimum degree at least $k$}.\hspace{2pt}\right]  \nonumber \\ &    \hspace{-2pt}=\hspace{-2pt} e^{- \frac{e^{-\lim_{n \to \infty}{\alpha_n}}}{(k-1)!}} \hspace{-2pt}=\hspace{-2pt}  \begin{cases} 0,&\text{\hspace{-8pt}if  }\lim_{n \to \infty}{\alpha_n} \hspace{-2pt}=\hspace{-2pt}- \infty, \\
1,&\text{\hspace{-8pt}if  }\lim_{n \to \infty}{\alpha_n} \hspace{-2pt}=\hspace{-2pt} \infty, \\
e^{- \frac{e^{-\alpha^{*}}}{(k-1)!}},&\text{\hspace{-8pt}if  }\lim_{n \to \infty}{\alpha_n} \hspace{-2pt}=\hspace{-2pt} \alpha^{*}\hspace{-2pt}\in\hspace{-2pt} (-\infty, \hspace{-1.5pt}\infty). \end{cases}\nonumber
 \end{align}
\end{thm}

\begin{thm}[\textbf{Perfect matching containment in binomial random $s$-intersection graphs \cite{2015-ISIT-s-intersection}}]  \label{pm:thm:bin}
For a binomial random $s$-intersection graph
$H_s(n,t_n,P_n)$, if there is a sequence $\beta_n$ with $\lim_{n \to \infty}{\beta_n} \in [-\infty, \infty]$
such that
\begin{align}
\frac{1}{s!} \cdot {t_n}^{2s}{P_n}^{s}  & = \frac{\ln  n   +
 {\beta_n}}{n}, \nonumber
\end{align}
then under $P_n  =  \Omega(n^c)$ for
a constant $c>2-\frac{1}{s}$, it holds that
 \begin{align}
& \lim_{n \to \infty}  \mathbb{P}[\hspace{2pt}H_s(n,t_n,P_n)\text{ has at least one perfect matching.}\hspace{2pt}] \nonumber \\ &   \hspace{-1pt}=\hspace{-1.5pt} e^{- e^{-\lim\limits_{n \to \infty}{\beta_n}}} \hspace{-1.5pt}=\hspace{-1.5pt}  \begin{cases} 0,&\text{\hspace{-4pt}if  }\lim_{n \to \infty}{\beta_n} \hspace{-1.5pt}=\hspace{-1.5pt}- \infty, \\
1,&\text{\hspace{-4pt}if  }\lim_{n \to \infty}{\beta_n} \hspace{-1.5pt}=\hspace{-1.5pt} \infty, \\
e^{-e^{- \beta^{*}}},&\text{\hspace{-4pt}if  }\lim_{n \to \infty}{\beta_n} \hspace{-1.5pt}=\hspace{-1.5pt} \beta^{*}\hspace{-1.5pt}\in\hspace{-1.5pt} (-\infty, \hspace{-1.5pt}\infty). \end{cases}\nonumber
 \end{align}
\end{thm}



\begin{thm}[\textbf{Hamilton cycle containment in binomial random $s$-intersection graphs \cite{2015-ISIT-s-intersection}}]  \label{hc:thm:bin}
For a binomial random $s$-intersection graph
$H_s(n,t_n,P_n)$, if there is a sequence $\gamma_n$ with $\lim_{n \to \infty}{\gamma_n} \in [-\infty, \infty]$
such that
\begin{align}
\frac{1}{s!} \cdot {t_n}^{2s}{P_n}^{s} & = \frac{\ln  n   +  \ln \ln n  +
 {\gamma_n}}{n}, \nonumber
\end{align}
then under $P_n  =  \Omega(n^c)$ for
a constant $c>2-\frac{1}{s}$, it holds that
 \begin{align}
& \lim_{n \to \infty}  \mathbb{P}[\hspace{2pt}H_s(n,t_n,P_n)\text{ has at least one Hamilton cycle.}\hspace{2pt}] \nonumber \\ &   \hspace{-1pt}=\hspace{-1.5pt} e^{- e^{-\lim\limits_{n \to \infty}{\gamma_n}}} \hspace{-1.5pt}=\hspace{-1.5pt}  \begin{cases} 0,&\text{\hspace{-4pt}if  }\lim_{n \to \infty}{\gamma_n} \hspace{-1.5pt}=\hspace{-1.5pt}- \infty, \\
1,&\text{\hspace{-4pt}if  }\lim_{n \to \infty}{\gamma_n} \hspace{-1.5pt}=\hspace{-1.5pt} \infty, \\
e^{-e^{- \gamma^{*}}},&\text{\hspace{-4pt}if  }\lim_{n \to \infty}{\gamma_n} \hspace{-1.5pt}=\hspace{-1.5pt} \gamma^{*}\hspace{-1.5pt}\in\hspace{-1.5pt} (-\infty, \hspace{-1.5pt}\infty). \end{cases}\nonumber
 \end{align}
\end{thm}


\begin{thm}[\textbf{$k$-Robustness in binomial random $s$-intersection graphs \cite{2015-ISIT-s-intersection}}]  \label{rb:thm:bin}
For a binomial random $s$-intersection graph
$H_s(n,t_n,P_n)$, if there is a sequence $\gamma_n$ with $\lim_{n \to \infty}{\gamma_n} \in [-\infty, \infty]$
such that
\begin{align}
\frac{1}{s!} \cdot {t_n}^{2s}{P_n}^{s} & = \frac{\ln  n   + {(k-1)} \ln \ln n  +
 {\gamma_n}}{n}, \nonumber
\end{align}
then under $P_n  =  \Omega(n^c)$ for
a constant $c>2-\frac{1}{s}$, it holds that
\begin{subnumcases}{ \hspace{-10pt}\lim\limits_{n \to \infty}\hspace{-1.5pt} \mathbb{P}[ H_s(n,t_n,P_n) \text{ is $k$-robust.}  ] \hspace{-1.5pt}=\hspace{-1.5pt}} 
\hspace{-3pt} 0, &\text{\hspace{-10pt}if $\gamma^*\hspace{-1.5pt}=\hspace{-1.5pt}-\infty$},  \label{RB-leq}
 \\  \hspace{-3pt}1,& \text{\hspace{-10pt}if $\gamma^*
\hspace{-1.5pt}=\hspace{-1.5pt}\infty$.}   \label{RB-geq}
 \end{subnumcases}
\end{thm}

\subsection{Results of random intersection graphs composed with Erd\H{o}s--R\'enyi graphs}  

\begin{thm}[\hspace{0pt}{\textbf{$k$-Connectivity in uniform random $1$-intersection graphs $\acap$
Erd\H{o}s--R\'enyi graphs
 by our work \cite{WiOpt15,ZhaoYaganGligor,ISIT}}}\hspace{0pt}]   \label{zhaothm}
 
 Consider a graph $G_1(n,K_n,P_n) \bcap \mathcal{G}_{ER}(n, q_n)$ induced by the composition of a uniform random $1$-intersection graph
$G_1(n,K_n,P_n)$ and an Erd\H{o}s--R\'enyi graph $\mathcal{G}_{ER}(n, q_n)$. With $s_n$ denoting the edge probability of $G_1(n,K_n,P_n) \bcap \mathcal{G}_{ER}(n, q_n)$, if there is a sequence $\alpha_n$ with $\lim_{n \to \infty}{\alpha_n} \in [-\infty, \infty]$
such that
\begin{align}
s_n & = \frac{\ln  n   + {(k-1)} \ln \ln n  +
 {\alpha_n}}{n}, \nonumber
\end{align}
then under $P_n = \Omega(n)$ and $\frac{K_n}{P_n} = o(1)$, it holds that
\begin{align}
&  \lim_{n \to \infty}\mathbb{P} \left[\hspace{2pt}G_1(n,K_n,P_n) \bcap \mathcal{G}_{ER}(n, q_n)\textrm{
is $k$-connected}.\hspace{2pt}\right] \nonumber \\
 & \hspace{-2pt}=\hspace{-2pt}  \lim_{n \to \infty} \mathbb{P}\bigg[\hspace{-4pt}\begin{array}{l}G_1(n,K_n,P_n) \bcap \mathcal{G}_{ER}(n, q_n)\\
 \hspace{-.2pt}\text{has a minimum degree at least
}k.\end{array}\hspace{-4pt}\bigg] \nonumber \\ &    \hspace{-2pt}=\hspace{-2pt} e^{- \frac{e^{-\lim_{n \to \infty}{\alpha_n}}}{(k-1)!}} \hspace{-2pt}=\hspace{-2pt}  \begin{cases} 0,&\text{\hspace{-8pt}if  }\lim_{n \to \infty}{\alpha_n} \hspace{-2pt}=\hspace{-2pt}- \infty, \\
1,&\text{\hspace{-8pt}if  }\lim_{n \to \infty}{\alpha_n} \hspace{-2pt}=\hspace{-2pt} \infty, \\
e^{-e^{- \alpha^{*}}},&\text{\hspace{-8pt}if  }\lim_{n \to \infty}{\alpha_n} \hspace{-2pt}=\hspace{-2pt} \alpha^{*}\hspace{-2pt}\in\hspace{-2pt} (-\infty, \hspace{-1.5pt}\infty). \end{cases}\nonumber
 \end{align}

\end{thm}

\begin{rem}

As summarized in Theorem \ref{zhaothm}, for $k$-connectivity in a uniform random $1$-intersection graph composed with an
Erd\H{o}s--R\'enyi graph,
our papers \cite{ZhaoYaganGligor,ISIT} show a zero--one law and later our another work \cite{WiOpt15} derives
 the asymptotically exact probability. For connectivity, Ya\u{g}an \cite{yagan_onoff} show a zero--one law under a weaker scaling.

\end{rem}

\begin{thm}[\hspace{0pt}{\textbf{$k$-Connectivity in uniform random $s$-intersection graphs $\acap$
Erd\H{o}s--R\'enyi graphs
 by our work \cite{JZISIT14}}}\hspace{0pt}]   
 
 Consider a graph $G_s(n,K_n,P_n) \bcap \mathcal{G}_{ER}(n, q_n)$ induced by the composition of a uniform random $s$-intersection graph
$G_s(n,K_n,P_n)$ and an Erd\H{o}s--R\'enyi graph $\mathcal{G}_{ER}(n, q_n)$. With $s_n$ denoting the edge probability of $G_s(n,K_n,P_n) \bcap \mathcal{G}_{ER}(n, q_n)$, if there is a sequence $\alpha_n$ with $\lim_{n \to \infty}{\alpha_n} \in [-\infty, \infty]$
such that
\begin{align}
s_n & = \frac{\ln  n   + {(k-1)} \ln \ln n  +
 {\alpha_n}}{n}, \nonumber
\end{align}
then under $P_n = \Omega(n)$ and $\frac{K_n}{P_n} = o(1)$, it holds that
\begin{align}
&  \hspace{-2pt}  \lim_{n \to \infty} \mathbb{P}\bigg[\hspace{-4pt}\begin{array}{l}G_s(n,K_n,P_n) \bcap \mathcal{G}_{ER}(n, q_n)\\
 \hspace{-.2pt}\text{has a minimum degree at least
}k.\end{array}\hspace{-4pt}\bigg] \nonumber \\ &    \hspace{-2pt}=\hspace{-2pt} e^{- \frac{e^{-\lim_{n \to \infty}{\alpha_n}}}{(k-1)!}} \hspace{-2pt}=\hspace{-2pt}  \begin{cases} 0,&\text{\hspace{-8pt}if  }\lim_{n \to \infty}{\alpha_n} \hspace{-2pt}=\hspace{-2pt}- \infty, \\
1,&\text{\hspace{-8pt}if  }\lim_{n \to \infty}{\alpha_n} \hspace{-2pt}=\hspace{-2pt} \infty, \\
e^{-e^{- \alpha^{*}}},&\text{\hspace{-8pt}if  }\lim_{n \to \infty}{\alpha_n} \hspace{-2pt}=\hspace{-2pt} \alpha^{*}\hspace{-2pt}\in\hspace{-2pt} (-\infty, \hspace{-1.5pt}\infty). \end{cases}\nonumber
 \end{align}

\end{thm}

\subsection{Results of random intersection graphs composed with random geometric graphs}  

\begin{thm}[\hspace{0pt}{\textbf{Connectivity in uniform random $1$-intersection graphs $\acap$
random geometric graphs without the boundary effect
 by our work \cite{ZhaoAllerton}}}\hspace{0pt}]   \label{thm-a1-bd}
 
 Consider a graph $G_1(n,K_n,P_n) \bcap \mathcal{G}_{RGG}(n, r_n, \mathcal{T})$ induced by the composition of a uniform random $s$-intersection graph
$G_s(n,K_n,P_n)$ and a random geometric graph $\mathcal{G}_{RGG}(n, r_n, \mathcal{T})$, where $\mathcal{T}$ is a torus of unit area. If 
\begin{align}
 \pi {r_n}^2 \cdot \frac{{K_n}^2}{P_n} & \sim a \cdot \frac{\ln n}{n} \label{thm:t:rnKnPn}
\end{align}
for some positive constant $a$, then under $K_n = \omega(\ln n)$,
$\frac{{K_n}^2}{P_n}  =  O\big(\frac{1}{\ln n}\big)$, $\frac{{K_n}^2}{P_n} = \omega\big(\frac{\ln
n }{n}\big)$, $\frac{K_n}{P_n} =
o\big(\frac{1}{n}\big)$, it holds that 
\begin{align}
 & \lim\limits_{n \to \infty}\mathbb{P}\left[ \hspace{2pt} \textrm{$G_1(n,K_n,P_n) \bcap \mathcal{G}_{RGG}(n, r_n, \mathcal{T})$}
 \textrm{ is connected.} \hspace{2pt}  \right] \nonumber \\  & \quad =
\begin{cases} 0, &\textrm{if $a<1$},   \\  1, &\textrm{if $a>1$}.\end{cases} \nonumber
\end{align}
\end{thm}

\begin{thm}[\hspace{0pt}{\textbf{Connectivity in uniform random $1$-intersection graphs $\acap$
random geometric graphs with the boundary effect
 by our work \cite{ZhaoAllerton}}}\hspace{0pt}]   \label{acc}
 
 Consider a graph $G_1(n,K_n,P_n) \bcap \mathcal{G}_{RGG}(n, r_n, \mathcal{S})$ induced by the composition of a uniform random $s$-intersection graph
$G_s(n,K_n,P_n)$ and a random geometric graph $\mathcal{G}_{RGG}(n, r_n, \mathcal{S})$, where $\mathcal{S}$ is a square of unit area. If 
\begin{subnumcases}  { \pi {r_n}^{2} \cdot \frac{{K_n}^2}{P_n} \sim}
\hspace{-2pt} \normalsize \selectfont  b \cdot \frac{\ln
\frac{n P_{n}}{{K_{n}}^2}}{n},   ~~~\textrm{\normalsize \selectfont for~} \normalsize \selectfont \frac{{K_n}^2}{P_n}  = \omega \bigg( \frac{1}{n^{1/3}\ln n} \bigg), \nonumber \\
\hspace{-2pt} \normalsize \selectfont b \cdot \frac{4\ln
\frac{P_n}{{K_n}^2} }{n}, ~\textrm{\normalsize \selectfont for~}
\normalsize \selectfont  \frac{{K_n}^2}{P_n}  = O \bigg( \frac{1}{n^{1/3}\ln n}
\bigg), \nonumber
 \end{subnumcases}
for some positive constant $b$, then under $K_n  \hspace{1pt}= \hspace{1pt} \omega(\ln n)$,
$\frac{{K_n}^2}{P_n}  =  O\big(\frac{1}{\ln n}\big)$, $\frac{{K_n}^2}{P_n} = \omega\big(\frac{\ln
n }{n}\big)$, $\frac{K_n}{P_n} =
o\big(\frac{1}{n}\big)$, it holds that 
\begin{align}
 & \lim\limits_{n \to \infty}\mathbb{P}\left[ \hspace{2pt} G_1(n,K_n,P_n) \bcap \mathcal{G}_{RGG}(n, r_n, \mathcal{S})
 \textrm{ is connected.}\hspace{2pt} \right]  \nonumber \\  & \quad =
\begin{cases} 0, &\textrm{if $b
<1$},   \\  1, &\textrm{if $b
>1$}.\end{cases} \nonumber
\end{align}
\end{thm}

\begin{rem}
For the graph $G_1(n,K_n,P_n) \bcap \mathcal{G}_{RGG}(n, r_n, \mathcal{S})$,
Krzywdzi\'{n}ski and Rybarczyk \cite{Krzywdzi} and
Krishnan \emph{et al.} \cite{ISIT_RKGRGG} also obtain connectivity
results, but their results are weaker than that in Theorem \ref{acc} above; see \cite[Section 
VIII]{ZhaoAllerton} for details. Furthermore,  Pishro-Nik \emph{et
al.} \cite{Pishro} and Yi \emph{et al.} \cite{4151628} investigate the absence of isolated nodes.
 
\end{rem}

\section{A Comparison between Random Intersection Graphs (resp., Their Intersections with Other Random Graphs) and Erd\H{o}s--R\'enyi Graphs} \label{compare}

To compare our studied graphs with Erd\H{o}s--R\'enyi graphs, we summarize below the results of Erd\H{o}s--R\'enyi graphs shown in prior work.

\begin{lem}[\hspace{0pt}{\textbf{$k$-Connectivity in Erd\H{o}s--R\'enyi graphs by \cite[Theorem 1]{erdos61conn}}}\hspace{0pt}] \label{lem:ER:PM}
For an Erd\H{o}s--R\'enyi graph $\mathcal{G}_{ER}(n,q_n)$, if there is a sequence $\alpha_n$ with $\lim_{n \to \infty}{\alpha_n} \in [-\infty, \infty]$
such that $q_n  = \frac{\ln  n   + (k-1) \ln \ln n
 {\alpha_n}}{n}$, 
 then it holds that
 \begin{align}
&  \lim_{n \to \infty}\mathbb{P} \left[\hspace{2pt}\mathcal{G}_{ER}(n,q_n)\textrm{
is $k$-connected}.\hspace{2pt}\right] \nonumber \\
 & \hspace{-2pt}=\hspace{-2pt}  \lim_{n \to \infty}\mathbb{P} \left[\hspace{2pt}\mathcal{G}_{ER}(n,q_n)\textrm{
has a minimum degree at least $k$}.\hspace{2pt}\right]  \nonumber \\ &    \hspace{-2pt}=\hspace{-2pt} e^{- \frac{e^{-\lim_{n \to \infty}{\alpha_n}}}{(k-1)!}} \hspace{-2pt}=\hspace{-2pt}  \begin{cases} 0,&\text{\hspace{-8pt}if  }\lim_{n \to \infty}{\alpha_n} \hspace{-2pt}=\hspace{-2pt}- \infty, \\
1,&\text{\hspace{-8pt}if  }\lim_{n \to \infty}{\alpha_n} \hspace{-2pt}=\hspace{-2pt} \infty, \\
e^{- \frac{e^{-\alpha^{*}}}{(k-1)!}},&\text{\hspace{-8pt}if  }\lim_{n \to \infty}{\alpha_n} \hspace{-2pt}=\hspace{-2pt} \alpha^{*}\hspace{-2pt}\in\hspace{-2pt} (-\infty, \hspace{-1.5pt}\infty). \end{cases}\nonumber
 \end{align}
 \end{lem}

\begin{lem}[\hspace{0pt}{\textbf{Perfect matching containment in Erd\H{o}s--R\'enyi graphs by \cite[Theorem 1]{erdosPF}}}\hspace{0pt}]  \label{lem:ER:PM}
For an Erd\H{o}s--R\'enyi graph $\mathcal{G}_{ER}(n,q_n)$, if there is a sequence $\beta_n$ with $\lim_{n \to \infty}{\beta_n} \in [-\infty, \infty]$
such that $q_n  = \frac{\ln  n   +
 {\beta_n}}{n}$, 
 then it holds that
 \begin{align}
 \lim_{n \to \infty}   \mathbb{P}[\hspace{2pt} \mathcal{G}_{ER}(n,q_n)\text{ has a perfect matching.} \hspace{2pt}] &  = e^{- e^{-\lim\limits_{n \to \infty}{\beta_n}}}. \nonumber
 \end{align}
 \end{lem}

\begin{lem}[\hspace{0pt}{\textbf{Hamilton cycle containment in Erd\H{o}s--R\'enyi graphs by \cite[Theorem 1]{erdosHC}}}\hspace{0pt}]    \label{lem:ER:HC}
For an Erd\H{o}s--R\'enyi graph $\mathcal{G}_{ER}(n,q_n)$, if there is a sequence $\gamma_n$ with $\lim_{n \to \infty}{\gamma_n} \in [-\infty, \infty]$
such that $q_n = \frac{\ln  n   + \ln \ln n +
 {\gamma_n}}{n}$,  
 then it holds that
 \begin{align}
 \lim_{n \to \infty}   \mathbb{P}[\hspace{2pt} \mathcal{G}_{ER}(n,q_n)\text{ has a Hamilton cycle.} \hspace{2pt}] &  = e^{- e^{-\lim\limits_{n \to \infty}{\gamma_n}}}. \nonumber
 \end{align}
 \end{lem} 
 
 \begin{lem}[\hspace{0pt}{\textbf{$k$-Robustness in Erd\H{o}s--R\'enyi graphs by \cite[Theorem 3]{6425841}
 and  \cite[Lemma 1]{ZhaoCDC}}}\hspace{0pt}]     
\label{lem:ER:RB} For an Erd\H{o}s--R\'{e}nyi graph
$\mathcal{G}_{ER}(n,q_n)$, with a sequence $\delta_n$ for all $n$ through \vspace{-2pt}
\begin{align}
q_n = & \frac{\ln  n + {(k-1)} \ln \ln n + {\delta_n}}{n}
 \label{hat_pn},   \vspace{-2pt} 
 \end{align}
then it holds that \vspace{-2pt}
\begin{align}
 & \lim_{n \to \infty} \hspace{-1pt}\mathbb{P} \big[\hspace{2pt}\mathcal{G}_{ER}(n,q_n)\hspace{2pt}\textrm{
is $k$-robust}.\big]  = \begin{cases} 0, \textrm{ if $\lim_{n \to
\infty}{\delta_n} \hspace{-2pt}=\hspace{-2pt}-\infty$}, \\  1,
\textrm{ if $\lim_{n \to \infty}{\delta_n}
\hspace{-2pt}=\hspace{-2pt}\infty$.}
\end{cases} \vspace{-2pt} \label{er_rb}
 \end{align}
\end{lem}

From Theorems \ref{thm-a1}--\ref{thm-a1-bd} and Lemmas \ref{lem:ER:PM}--\ref{lem:ER:RB}, random graphs  $G_1(n,K_n,P_n)$, $G_s(n,K_n,P_n)$, $H_1(n,t_n,P_n)$, $H_s(n,t_n,P_n)$, $G_1(n,K_n,P_n) \bcap \mathcal{G}_{ER}(n, q_n)$, $G_s(n,K_n,P_n) \bcap \mathcal{G}_{ER}(n, q_n)$, and $G_1(n,K_n,P_n) \bcap \mathcal{G}_{RGG}(n, r_n, \mathcal{T})$ under the conditions in the respective theorems have threshold behaviors for the respective properties similar to Erd\H{o}s--R\'{e}nyi graphs with the same edge probabilities. However, these graphs may be different from Erd\H{o}s--R\'{e}nyi graphs under other conditions or for other properties; e.g., $G_1(n,K_n,P_n)$ is shown to be more clustered than an Erd\H{o}s--R\'{e}nyi graph with the same edge probability \cite{YaganTriangle}.


\section{Conclusion}
\label{sec:Conclusion}

Random intersection graphs have recently been studied in the literature extensively and used in diverse applications. In this paper, we summarize results of random intersection graphs and their compositions with other random graphs, mostly from our prior work. We also discuss the applications of random intersection graphs to secure wireless communication and social networks.

\end{document}